\documentclass[twocolumn,amssymb,aps,nofootinbib,floatfix]{revtex4}
\usepackage{epsfig}
\newcommand{\be}{\begin{eqnarray}}
\newcommand{\ee}{\end{eqnarray}}
\newcommand{\ave}[1]{\left\langle #1 \right\rangle}

\newcommand{\order}[1]{ \mathcal{O} \left( #1 \right) }

\newcommand{\tilro}{\tilde{\rho}}

\begin{document} \hbadness=10000
\topmargin -0.8cm\oddsidemargin = -0.7cm\evensidemargin = -0.7cm
\title{Flavoring the Quark-Gluon Plasma with Charm}
\author{Giorgio Torrieri$^{1}$} 
\email{torrieri@th.physik.uni-frankfurt.de}
\author{Jorge Noronha$^2$}
\email{noronha@phys.columbia.edu}
\affiliation{
$^1$FIAS,
  J.W. Goethe Universit\"at, Frankfurt A.M., Germany  \\
$^2$Department of
Physics, Columbia University, 538 West 120$^{th}$ Street, New York,
NY 10027, USA}
\date{March 29, 2010}  

\begin{abstract}
We demonstrate that a Quark-Gluon Plasma (QGP) with a dilute admixture of heavy quarks has, in general, a lower speed of sound than a ``pure'' QGP without effects from heavy flavors.  The change in the speed of sound is sensitive to the details of the theory, making the hydrodynamic response to ``flavoring'' a sensitive probe of the underlying microscopic dynamics. We suggest that this effect may be measured in ultrarelativistic heavy ion collisions by relating the event-by-event number of charm quarks to flow observables such as the average transverse momentum. 
\end{abstract}
\maketitle
One of the most widely cited findings in ultrarelativistic heavy ion collisions concerns the discovery of a
``perfect liquid'' in collisions at the Relativistic Heavy Ion Collider (RHIC) 
\cite{v2popular,whitebrahms,whitephobos,whitestar,whitephenix}. 
The evidence for these claims comes from the successful modeling of the anisotropic
expansion of the matter in the early stage of the reaction by means of ideal
 hydrodynamics \cite{heinz,shuryak,huovinen}, as well as the presence of jet-flow correlations that exhibit a conical pattern \cite{mach3,mach4,mach5,mach6,mach7,mach8,mach1,mach2,mach9}.

A considerable amount of study has been carried out to understand why the transport properties of the system created at RHIC are so different (the average $\eta/s$,shear viscosity to entropy density ratio is so much lower) than the predicted properties of a weakly coupled QGP \cite{etaswqgp}.   This question is currently unsettled, with the observed liquid being either described as a
strongly interacting (t'Hooft coupling constant $\lambda \gg 1$) QGP \cite{sqgp}, a bound
state QGP \cite{bsqgp}, a (turbulent) Glasma \cite{glasma} with instabilities, a thermalizing Hagedorn resonance gas \cite{jaki}, or a ``semi-QGP'' with Polyakov loops as active degrees of freedom \cite{robhidaka}.  

Less attention has gone into using hydrodynamics as a tool to link QGP phenomenology to observables known from first principle quantum chromodynamics (QCD), notably the equation of state above deconfinement. It should be stressed that studies of collective flow were among the earliest predicted observables to probe the thermal properties of heated and
compressed nuclear matter \cite{Scheid:1974yi}. As the transverse flow is connected
to the pressure gradients in the early stage of the reaction, it provides information on the equation 
of state (EoS) and might therefore be used to search for abnormal matter states and 
phase transitions \cite{Stoecker:1979mj,Hofmann:1976dy,Stoecker:1986ci}.

The advent of the Large Hadron Collider (LHC) gives us a qualitatively new laboratory for the study of ``soft flow'' observables.   A naive extrapolation, based on the boost-invariance assumption, logarithmic scaling of multiplicity with respect to center of mass energy, and an ideal gas equation of state, predicts that the initial temperature of the LHC will be 1.5-3 times the initial temperature of RHIC, $500-1000$ MeV.   Thus, one expects that the soft properties of the two regimes are largely the same, or at least comparable. However, statistics for {\em rare probes} produced in hard initial interactions (e.g., jets, heavy quarks) should be better by at least an order of magnitude and could be used to qualitatively Gauge the thermal and statistical properties of the heavy ion background.



In this work we suggest a new way to correlate rare, heavy quark probes to the thermodynamic properties of the QGP: the change of the equation of state due to the presence of a dilute admixture of heavy quarks. We show that the speed of sound in a heavy ``flavored'' QGP decreases with the (small) concentration of charm quarks produced in the early stages of the collisions. This leads to a novel anti-correlation between the average transverse momentum, $\langle p_T \rangle $, and the total number of charm quarks, $N_{charm}$, that could be measured at the LHC.      

In general, charm is not expected to be {\em chemically} equilibrated in heavy ion collisions.   The bulk of charm content should be produced by ``hard'' processes in the initial state \cite{ncc1,ncc2,ramona,lhcpred} at a concentration far above their equilibrium expectation \cite{janc}.   The abundance of $c \overline{c}$ pairs produced in heavy ion collisions at the LHC is expected to be reasonably high ($\sim 10^{1-2}$, of course parametrically smaller than the total $\sim 10^{4}$ multiplicity), and it follows from the strongly coupled nature of the bulk medium that they might achieve {\em thermal} equilibrium \cite{teaneyc,rapp,janc,pbm}.  
Thus, heavy quarks might well function as a ``dilute flavor'' capable of modifying the equation of state of the system by a calculable amount, analogously to the way a dilute admixture of salt modifies the heat capacity of water.


Provided the total charm abundance is not changed between production and freeze-out (a reasonable assumption in the dilute limit) and charm is thermally equilibrated, the dimensionless quantity $\tilro = \rho/s \sim \rho T^{-3} \ll 1$, where $\rho$ is the charm number density and $s$ is the entropy density is a useful parameter to describe the heavy flavorness of the event (note that $\rho_c = \rho_{\overline{c}}=\rho/2$). If the charm distribution is approximately homogeneous, $\tilro$ can be assumed to be conserved throughout the hydrodynamic stage provided that charm diffusion is negligible (which seems to be the case at RHIC energies \cite{teaneyc,rapp}).  Since entropy $s$ is, to a good approximation four times the multiplicity \cite{Fer50}, $\tilro$ can be {\em measured} event-by-event if the detector has appropriate charm reconstruction.   Provided an appropriately large sample of events is available, the experimentalist can therefore select a sample of arbitrarily high $\tilro$ ``flavorful'' events and study the effect of flavorness on the equation of state, analogously to the way 
the effect of impurities on a solution's equation of state can be studied by adding the required amount of impurities by hand in the lab.

The effects of charm quarks on the QGP EoS have been studied within perturbation theory \cite{Laine:2006cp}, on the lattice \cite{latticesalt,detar}, and also within recently derived sum rules \cite{chernodub}. In the dilute limit relevant for heavy ion collisions, $\tilro \ll 1$, the effects of charm quarks on the pressure of an interacting plasma composed of up, down, and strange quarks can be calculated to $\mathcal{O}\left(T/M_{quark} \right)$ (small at RHIC, and most likely at the LHC) by adding a Polyakov loop density to the free energy density

\begin{equation}
\mathcal{F}(T) = \mathcal{F}_0(T) +\tilro \,s_0(T) \, F_Q (T)
\label{noronhaeq1}
\end{equation} 
where $F_Q(T) = -T \ln \ell(T)$ and $\ell(T)$ is the renormalized Polyakov loop (obtainable from lattice calculations) and the quantities underscored with 0 denote the values before the charm flavor was included (e.g., $s_0$ is the lattice data for the entropy density of a 2+1 QGP \cite{bazavov}). Corrections to the free energy will generally come with higher powers of $\tilro$ and can, thus, be safely assumed to be small in the dilute approximation, as well as finite mass effects which again are negligible for most if not all of the hydrodynamic stage. The density of Polyakov loops considered here describes how much the thermodynamics of the system changes by the addition of the heavy flavor. 


This simple model predicts that the speed of sound of the charmed QGP, $c_s(T)$, is lower than the value found in the standard 2+1 QGP, $c_{s\,0}(T)$, near the crossover phase transition. This can be seen as follows. The Polyakov loop measures the excess in free energy due to the addition of an infinitely massive source of fundamental flux (see \cite{jorgef} for a discussion of the physical interpretation and general properties of Polyakov loops in QCD and in holographic Gauge theories). From Eq.\ (\ref{noronhaeq1}) one can show that the entropy density of the charmed QGP is   

\begin{equation}
\ s(T) = s_0(T) - \tilro \,\frac{s_0(T)}{T}\left[U_{Q}(T)+\frac{F_Q(T)}{c_{s\,0}^2}(1-c_{s\,0}^2)   \right]
\label{noronhaeq2}
\end{equation}   
where $U_Q\equiv F_Q -T \,dF_Q/dT$ (note that $\tilro$ is temperature independent). In confining theories that are also asymptotically free, near the deconfinement phase transition both $U_Q$ and $F_Q$ are positive \cite{jorgef}. Thus, the small concentration of charm quarks added to the system introduces correlations that decrease the entropy density of the system. The speed of sound is generally related to the entropy density $s$ in thermodynamic equilibrium as follows 
\begin{equation}
c_s^2 = \frac{d\ln T}{d\ln s}\,.
\end{equation}
Since both $U_Q$ and $F_{Q}$ become negative far from the phase transition in QCD \cite{jorgef}, it is easy to see that the specific heat of the charmed QGP increases with $\tilro$, which then implies that the speed of sound $c_s (\tilro \neq 0,T) < c_{s\,0}(T)$.   All this has a simple physical explanation:  Correlations between the medium and slowly moving heavy quarks contribute to the energy density but not to pressure, thus lowering the system's response to perturbations. Note that its the {\em correlations}, rather than the quarks, assumed here to be infinitely heavy, that do this.

\begin{figure}[t]
\epsfig{width=8cm,clip=,figure=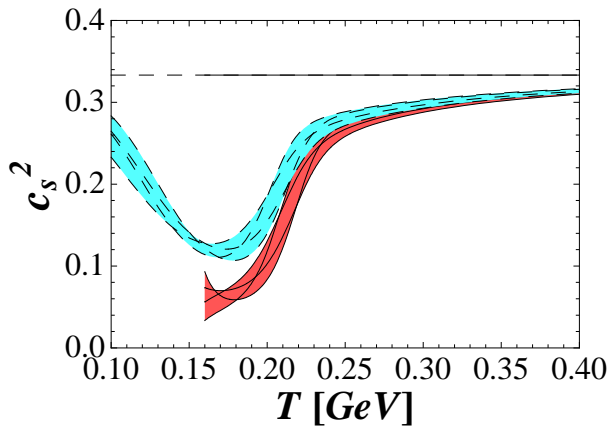}
\caption{(color online) \label{saltcs} The speed of sound dependence on the charm quark concentration in the QGP, for $\tilro=0$ (cyan, dashed lines) and $\tilro=0.1$ (red, solid lines).  The width of the bands denotes lattice uncertainties \cite{bazavov} in the speed of sound of a 2+1 QGP, while the thick line denotes the conformal, non-interacting value where $c_s^2=1/3$. }
\end{figure}

Fig. \ref{saltcs} shows our estimate for the speed of sound derived within the phenomenological model in Eq.\ (\ref{noronhaeq1}) using the expectation value of the Polyakov loop extracted from the lattice  (2+1 QGP with almost physical quark masses \cite{bazavov}). One can see that the main effect comes from the region near the phase transition (where there is a minimum in the speed of sound) but well before the Polyakov loop expectation value reaches its asymptotic high-T limit, leading to a negative shift of the speed of sound from its value in a 2+1 QGP (as computed on the lattice \cite{bazavov}).

An analysis of the recent lattice simulations of a 2+1+1 QGP \cite{detar} validates these conclusions, but the effect we are predicting is parametrically larger since it is implicitly assumed in Ref.\ \cite{detar} that $\tilro$ is given by its equilibrium value. In heavy ion collisions, initial state interactions can bring $\tilro$ up from the equilibrium value by two orders of magnitude \cite{janc,pbm,ramona}.   If charm is thermalized and dilute, it is reasonable to expect the effect will be appropriately amplified.

Our estimate stops close to $T_c$, as $-T \ln \ell(T) \rightarrow \infty$ in the confining phase (where the Polyakov loop expectation value vanishes).    Mathematically, one can trust our approach as long as the heavy quark is much heavier than any other scale in the system, ie $-T \ln \ell(T) \ll M_q$.  At some point in the approach to confinement, however, this approximation breaks down and the Polyakov loop method becomes unreliable.  Physically, in a confining theory with light quarks, string breaking in the confined phase means that the free energy of the system is not anymore related to the Polyakov loop expectation value.

To estimate the contribution of flavoring in the confined phase, we make the reasonable assumption that, just like a flavorless confined QCD thermal system \cite{philipsen},  flavorful confined QCD is well-described by the hadron resonance gas model.    In this case, flavoring can be approximated by an admixture of heavy mesons in a gas of pions.  The latter has a speed of sound of $c_s^{light} \simeq 1/\sqrt{3}$ (ultra-relativistic ideal gas), while the former will have a speed of sound of $c_s^{heavy} \simeq \sqrt{5T/(3M_{meson})}$ (non-relativistic).   It is not difficoult to see that the speed of sound of the mixture \cite{landau} will go as
\begin{equation}
c_s^2 \sim \frac{1}{3} - \order{\tilro \frac{T}{M_{meson}}}
\end{equation}
parametrically smaller than the contribution in the deconfined phase, which is just $\order{\tilro}$.    Thus, the flavoring effect on the speed of sound is
{\em specific to the deconfined phase}, and it is driven by the {\em correlation} between the quark and the medium due to strong QCD fields, rather than just by the large mass of the charm quark. 

The effect of flavoring on deconfined strongly coupled QCD is particularly interesting phenomenologically as it crucially depends on {\em both} the confining and asymptotically free nature of QCD. In fact, we expect the medium's response to heavy flavorness to be somewhat sensitive to the nature of the fixed point present in the ultraviolet (UV). According to the general arguments presented in \cite{jorgef}, in confining gauge theories with a trivial fixed point in the UV (such as QCD) $d\ell/dT$ is positive near the transition but it changes sign at higher temperatures and, thus, the Polyakov loop reaches its asymptotic value at high $T$ from above. However, in confining gauge theories with a holographic description in terms of supergravity $d\ell /dT \geq 0$ (this derivative can only vanish at the nontrivial fixed point in the UV). Therefore, the regularized Polyakov loop in QCD displays a bump at a given value of the temperature in the deconfined phase while in gauge theories described via supergravity the loop is a monotonically function of $T$ that reaches its asymptotic value at the fixed point from below \cite{jorgef}.  

\begin{figure*}[t]
\epsfig{width=8cm,clip=,figure=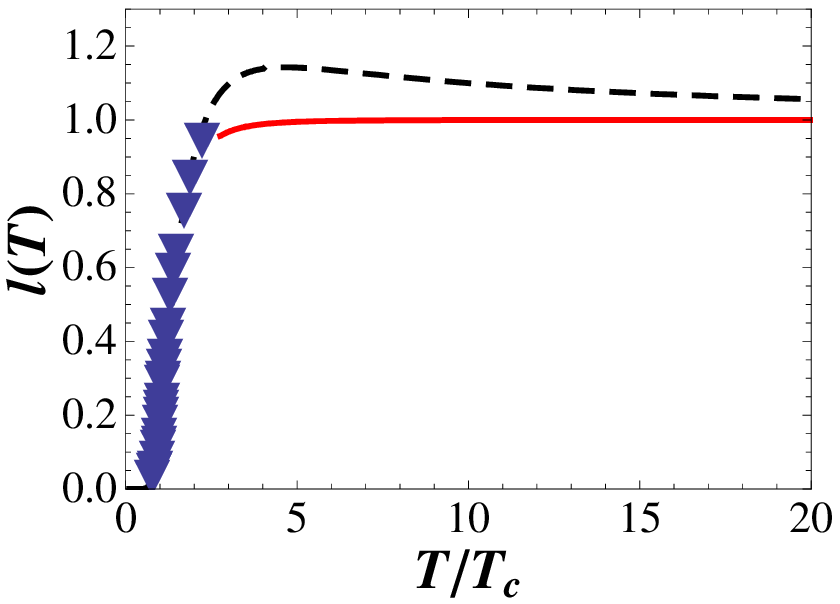}
\epsfig{width=8cm,clip=,figure=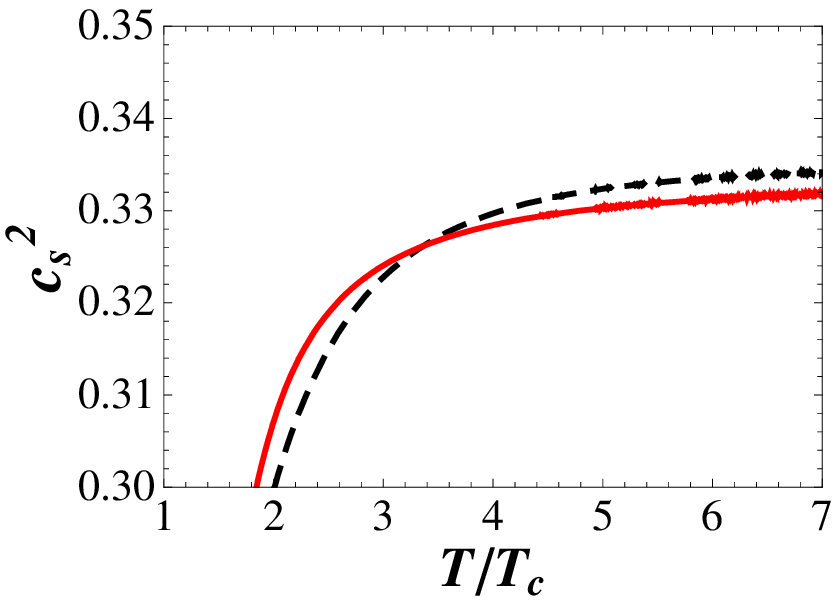}
\caption{(color online) \label{saltpol} 
Left panel: High temperature extrapolations of the Polyakov loop expectation value, each based on functions (see Eqs.(\ref{fit1} and (\ref{fit2})) that fit lattice data (blue triangles) \cite{bazavov} at low temperatures but have a very different high temperature behavior, one (black-dashed) compatible with asymptotic freedom, the other expected in a gauge theory described by a classical supergravity dual.  The corresponding speed of sound, with $\tilro=0.1$, are plotted on the right panel. }
\end{figure*}

To illustrate the difference between a theory that is non-interacting in the UV and a theory that has a nontrivial UV fixed point, we fitted the Polyakov loop data \cite{bazavov} (left panel of Fig. \ref{saltpol}) with the function
\begin{equation}
f_1 (x) =  \exp \left(\frac{a_1}{x}-\frac{b_1}{x^2} \right)
\label{fit1}
\end{equation}
where $x=T/T_c$, $T_c = 0.185$ GeV, and $a_1=1.24$ and $b_1=2.89$. This function, when extrapolated to high $T$, gives an evolution close to what is expected for QCD with a peak above $T_c$ \cite{jorgef}. We also considered another function that is qualitatively consistent with the evolution expected for a theory with confinement but with a classical supergravity description until $T \rightarrow \infty$ (i.e, no asymptotic freedom)
\begin{equation}
f_2 (x) =  \exp\left(-\frac{a_2}{x^b_2} \right)
\label{fit2}
\end{equation}
where $a_2=1.75$ and $b_2=3.62$. Note from the left panel of Fig.\ \ref{saltpol} that both functions fit the data equally well, but have a different qualitative behavior when extrapolated to higher temperatures.  As the right panel of Fig.\ \ref{saltpol} shows, the presence of a peak in $\ell$ produces a qualitatively different (though quantitatively small) modification of the speed of sound.

Thus, we expect the medium's response to heavy flavor in strongly coupled theories with supergravity duals to be, in general, qualitatively different from QCD. We shall further elaborate this point. For the $\mathcal{N}=4$ SYM theory \cite{maldacena} the modification of the free energy in presence of an infinitely heavy quark can be calculated by describing the Polyakov loop as a string going from the boundary to the black brane horizon.  In this picture $F_Q(T) = -\frac{1}{2} \sqrt{\lambda} T$.   Putting this and the equation of state for the $\mathcal{N}_{SYM}=4$ theory \cite{maldacena} in Eq. \ref{noronhaeq1} one can show that in this approximation $c_s^2=1/3$ regardless of the value of $\tilro$ and $\lambda$, as can be expected for an infinitely massive quark in a conformally invariant plasma.

More realistically, matter fields in the fundamental representation of $SU(N_c)$ can be introduced into the $\mathcal{N}=4$ SYM theory using D-branes in the gravitational description \cite{Karch:2002sh}. The effects of heavy quarks on the thermodynamics were extensively discussed in \cite{mateos} using the D3-D7 brane setup. The inclusion of heavy quarks in the plasma changes the speed of sound and, just like QCD, leads to a negative speed of sound dependence on the heavy quark content. The change in the speed of sound squared from 1/3 can be naively expected to be  $\sim -\order{\tilro \frac{T \sqrt{\lambda}}{ M_Q}}$. 

The advantage of this approach is that one has full control of not just the effects of heavy quarks on the equation of state but also on the transport coefficients.  In this model \cite{myers} $\eta/s=1/4\pi$ independently of $\tilro$ \cite{myers}, though both $\eta$ and $s$ get shifted (this is likely to be qualitatively different from physical QCD, for a discussion of the dependence of the shear viscosity on the Polyakov loop in a weakly coupled framework see \cite{robhidaka}).   

These effects produce, at least in principle, observable consequences that we will now explore. Note that $\tilro$ can be easily related to the number of charm quarks per unit of rapidity $dN_{charm}/dy$ and the charged multiplicity rapidity density $dN_{ch}/dy$ (due to the relationaship between entropy and multiplicity, \cite{Fer50}).  The first can be calculated via perturbative QCD \cite{ramona} and the second is expected to be logarithmic \cite{busza}.  Thus, using Bjorken's formulae \cite{bjorken} 
\begin{equation}
\tilro = \frac{1}{6}\frac{dN_{charm}/dy}{dN_{ch}/dy} \simeq \frac{1}{3} \frac{N_{coll}}{N_{part}} \frac{\sigma_{pp \rightarrow c \overline{c}}(\sqrt{s})}{A_0 \Delta y \ln \left( \frac{\sqrt{s}}{E_0} \right)}
\label{tilderho}
\end{equation}
Using the cross-section shown in \cite{ramona}, $N_{coll} \sim N_{part}^{4/3}$, and the parametrization for multiplicity in \cite{busza} ($A_0=1.14\times \pi (0.6 fm)^2,E_0=1.41 $ GeV), we estimate $\tilro$ to be small but non-negligible at top LHC energies (see Fig. \ref{tilderho}).

\begin{figure*}[t]
\epsfig{width=12cm,clip=,figure=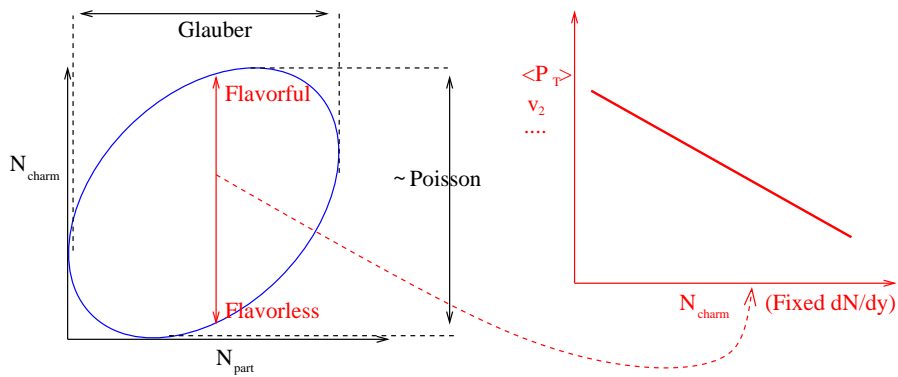}
\caption{(color online) \label{saltthermo} Left panel: The expected event-by-event distribution of events with respect to charm content and $N_{part}$, together with the cut required to analyse the response of the system to charm.  Rightpanel: The expected dependence of $\ave{p_T}$ with charm number.  $v_2$ is expected to have a similar dependence. }.
\end{figure*}

Of course, this estimate is illustrative only, due to the order of magnitude uncertainty in current pQCD calculations, as well as the theoretical controversy over $dN/dy$ \cite{lhcpred}.   Moreover, $c \overline{c}$ and (to a lesser extent) $dN/dy$ will vary event by event.    The fluctuation in the number of charm quarks is then expected to be Poissonian for a high enough event sample, while the fluctuation in $dN/dy$ is generally expected to follow KNO scaling \cite{kno}.  Hence, $\tilro$ is expected to vary considerably event-by-event, a fluctuation that, as usual, increases for smaller system sizes. The crucial issue, though, is that provided that charm can be reasonably reconstructed and there is a large enough event sample, $\tilro$ is an {\em experimental observable} capable of serving as a binning class for events (see Fig. \ref{saltthermo}).

As is well known, there is a connection between the speed of sound and the limiting average velocity of a hydrodynamic expansion with shock-like initial conditions, 
\begin{equation}
\label{vtprop}
\ave{\gamma_T v_T}_{freezeout} \sim f(N_{part}) \ave{c_s}_{\tau}
\end{equation}
where ``freezeout'' implies averaging over the freeze-out hypersurface while the subscript $\tau$ means the average is done over the hydrodynamic evolution. 
For a shallow shock this result is exact \cite{dirkhydro1}. 
While knowledge of the initial geometry is needed to establish the form of $f(N_{part})$, model calculations \cite{shuryak,hydro2,hydro3,hydro4} indicate that the dependence is not washed away even in steeper shocks and more complicated initial geometries. 

  The final transverse flow is in return connected to the average transverse momentum\footnote{Of course, by this we mean ``soft'' transverse momentum, with a $\sim 1$ GeV cut.  We underline this point as to exclude the rather high-momentum charm decay products. }
\begin{equation}
\label{ptvt}
\ave{p_T} \simeq T + m \ave{\gamma_T v_T}
\end{equation} 
In the past \cite{vanhove}, this was proposed as a signature of the mixed phase.  Since transverse flow, unlike elliptic flow, receives approximately equal contributions from all stages of the hydrodynamic expansion \cite{gyulflow}, the decrease of the speed of sound close to $T_c$ (Fig. \ref{saltcs}) could lower $\ave{p_T}$ for more charmed events with respect to charmless ones.  The coefficient associated with this heavy flavoring effect would be straightforwardly related to non-perturbative QCD via Eqs.\ (\ref{vtprop}) and (\ref{ptvt}), and Fig.\ \ref{saltcs}.   This effect might be easier to measure in smaller systems (such as $pp$ collisions if they also experience hydrodynamic flow at high energies \cite{ppflow1,ppflow2}) due to the greater event-by-event variation in $\tilro$ and more precise charm tracking.
 
\begin{figure}[t]
\epsfig{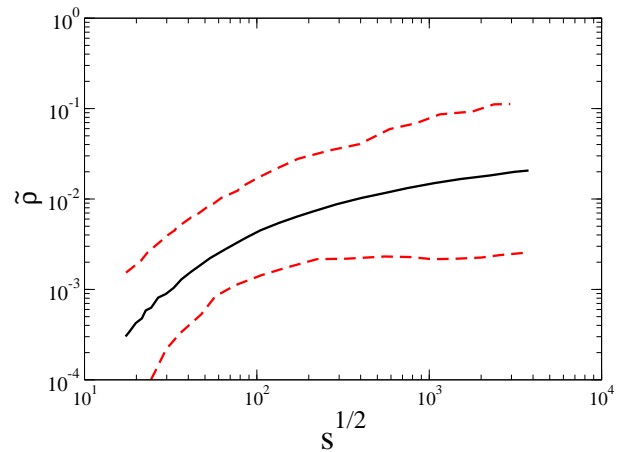}
\caption{(color online) \label{tilderho} $\tilro$ vs $\sqrt{s}$ for Pb-Pb collisions computed using Eq.\ (\ref{tilderho}).  The dashed lines denote the uncertainties in the perturbative QCD calculation \cite{ramona}.}
\end{figure}

The main requirement of such an analysis is the ability to experimentally gauge both the charm quark abundance and $\ave{p_T}$ {\em event-by-event}. At RHIC, event-by-event charm detection is nearly impossible, since heavy flavored particles are reconstructed only from the leptonic decay modes (``non-photonic electrons''), and this branching ratio only captures 70$\%$ of total decays.    At the LHC, however, it will be possible to find most charm particles in each event using primary vertex cuts \cite{charmalice,charmalice2}.

Another obstacle for detecting the admixture's effect on the QGP's thermal properties is cross-correlation. The average number of charmed particles is positively correlated to the average global multiplicity of the event. However, $\ave{p_T}$ is also correlated to this multiplicity \cite{huovinen} due to initial shock depth (the proportionality constant in Eq. \ref{vtprop} depends on $N_{part}$).  
It should be noted, however, that the effect of $\tilro$ on $\ave{p_T}$ is {\em opposite} to the correlation with $N_{part}$, events with a greater $\tilro$ should have a lower speed of sound and hence lower $\ave{p_T}$.
Hence, it should be possible to disentangle this charming effect from the flow response to fluctuations in the initial geometry.
Hence, and considering that  $N_{charm} \sim N_{collisions}$ rather than $N_{participants}$, this obstacle is {\em not} unsurmountable {\em provided} that the event-by-event charm tagging is precise enough. In fact, binning tightly in $dN/dy$ and looking for a correlation between $\ave{p_T}$ and $N_{charm}$ should separate the $\ave{p_T}$ correlation with $N_{part}$ from the heavy ``flavoring'' anti-correlation via the lower speed of sound.

A possible effect which would give correlations in the same sense as the effect proposed here is energy conservation (roughly, charm quarks need a lot of energy to be created, and that lowers $\ave{p_T}$).   This effect should however be suppressed by factorization and boost-invariance:   Charm quarks are created from partons with larger Bjorken $x$ than the bulk of soft particles at mid-rapidity, so the energy they take up comes from regions well away from mid-rapidity.  If boost-invariant hydrodynamics is the correct picture ( diffusion and convection across rapidity is negligible) we do not expect that energy conservation will lower $\ave{p_T}$ at mid-rapidity.

While a decrease in $\ave{p_T}$ with $N_{charm}$ could be straightforward to measure, it is not the only effect measurable by binning events on charm content. Since $\eta/s$ generally changes with charm content in a Plasma with Polyakov loops \cite{robhidaka}
$v_2/\epsilon$, the ratio of elliptic flow to initial eccentricity might depend on the amount of charm in the event.    Such measurements could be the key to distinguishing between weakly coupled approaches (where $\eta/s$ is allowed to vary) and strongly-coupled theories with gauge-gravity duals (where $\eta/s$ is fixed to leading order \cite{myers}).

With very high statistics, this effect could also be seen as a variation of the Mach cone angle \cite{mach3,mach4,mach5,mach6,mach7,mach8,mach1,mach2,mach9}
 and signal height in samples of events where the charm quarks are over or under-abundant.  

Finally, hotspots could lead to over-flavored regions which would stay together during the hydrodynamic stage, leading to a higher transverse momentum {\em fluctuations} $\ave{(\Delta p_T)^2}$, and a more anisotropic transverse momentum distribution, for more flavorful events. In this scenario $\ave{p_T}$ when binned by azimuthal angle would be correlated to charm direction.

In conclusion, we have described the modification of the thermal properties (the speed of sound and viscosity) due to the ``heavy flavoring'' of the plasma by an admixture of charm quarks. We have described how this effect might be measured in very high energy (LHC and higher energies) heavy ion collisions, where a sufficient amount of heavy flavor might be produced in the initial state to ``flavor'' the plasma to a level where the change in the equation of state will be observable.

We would like to thank C.~Nattrass and E.~Bruna for elucidating the experimental charm measurement capabilities at the LHC. We also thank O.~Linnyk, M.~Gyulassy, M.~Cheng, I.Mishustin,C.Greiner,D.Rischke and W.~Zajc for discussions. G.T. was (financially) supported by the Helmholtz International Center for FAIR within the framework of the LOEWE program (Landesoffensive zur Entwicklung Wissenschaftlich-Ökonomischer Exzellenz) launched by the State of Hesse. J.N. is supported by the US-DOE Nuclear Science Grant No.\ DE-FG02-93ER40764. G.~T. thanks M.~Gyulassy and Columbia University for the hospitality provided when this work was done.

\end{document}